# Formation of Graphene atop a Si adlayer on the C-face of SiC

Jun Li,[1] Qingxiao Wang,[2] Guowei He,[1] Michael Widom,[1] Lydia Nemec,[3,*] Volker Blum,[4] Moon Kim,[2] Patrick Rinke,[3,5] and Randall M. Feenstra[1,†]

[1]Dept. Physics, Carnegie Mellon University, Pittsburgh, PA 15213, USA

[2]Dept. Materials Science and Engineering, The University of Texas at Dallas, Richardson, TX 75080, USA

[3]Fritz-Haber-Institut der Max-Planck-Gesellschaft, D-14195 Berlin, Germany

[4]Dept. Mechanical Engineering and Materials Science, Duke University, Durham, NC 27708 USA

[5]Dept. Applied Physics, Aalto University, P.O. Box 11100, Aalto FI-00076, Finland

## Abstract

The structure of the SiC($000\bar{1}$) surface, the C-face of the {0001} SiC surfaces, is studied as a function of temperature and of pressure in a gaseous environment of disilane ($Si_2H_6$). Various surface reconstructions are observed, both with and without the presence of an overlying graphene layer (which spontaneously forms at sufficiently high temperatures). Based on cross-sectional scanning transmission electron microscopy measurements, the interface structure that forms in the presence of the graphene is found to contain $1.4-1.7$ monolayers (ML) of Si, a somewhat counter-intuitive result since, when the graphene forms, the system is actually under C-rich conditions. Using *ab initio* thermodynamics, it is demonstrated that there exists a class of Si-rich surfaces containing about 1.3 ML of Si that are stable on the surface (even under C-rich conditions) at temperatures above ~400 K. The structures that thus form consist of Si adatoms atop a Si adlayer on the C-face of SiC, with or without the presence of overlying graphene.

## I. Introduction

Formation of graphene on SiC, by heating the SiC and producing preferential sublimation of Si compared to C, has been studied extensively for more than a decade.[1] The (0001) surface, known as the *Si-face* of the two types of {0001} surfaces, has been employed in most of those studies; graphene with good structural and electronic properties can be produced on that surface.[2,3,4] It is known that between the graphene and the SiC there is an intermediate layer, a so-called buffer layer, consisting of a graphene-like structure but with some bonding to the underlying SiC, forming a $(6\sqrt{3}\times6\sqrt{3})$-*R*30° unit cell.[5] As additional Si is sublimated from the SiC, this buffer layer eventually converts to pristine graphene and a new buffer layer forms below it.[6,7,8,9] Additionally, the buffer layer can be decoupled from the SiC by introduction of hydrogen or oxygen.[3,10,11,12]

For graphene formation on the $(000\bar{1})$ surface of SiC, known as the *C-face*, the situation is found to be more complex than for the Si-face; there appears to be more than one way to form graphene on the surface (since various reconstructions are found at the interface),[13,14] and the structural quality of the graphene on the C-face is generally worse than for the Si-face.[2,15,16,17] However, considerable improvement in the quality of graphene on the C-face is achieved by performing the growth in a confined space, either utilizing "confinement controlled sublimation (CCS)" in a small, nearly closed carbon ampoule,[18,19,20,21] or simply by stacking two SiC wafers together (related methodologies have also been used for improvements in graphene quality on the Si-face,[2,16,17] although without any fundamental change in interface structure in that case). In these confined geometries, presumably the Si partial pressure above the SiC surface is much higher in these situations than in vacuum, i.e. a situation closer to thermodynamic equilibrium is achieved.[9,18] Similar improvement in graphene quality is also found when the formation is performed under an applied pressure of disilane (Si$_2$H$_6$) gas of $P_d \approx 10^{-5}$ Torr.[14,22,23,24] Exceptional results for the electronic properties of the C-face graphene have been obtained for samples formed in the confined geometry.[18,25]

---

* Present address: Carl Zeiss AG Digital Innovation Partners, Kistlerhofstraße 70, 81379 Munich, Germany

† feenstra@andrew.cmu.edu

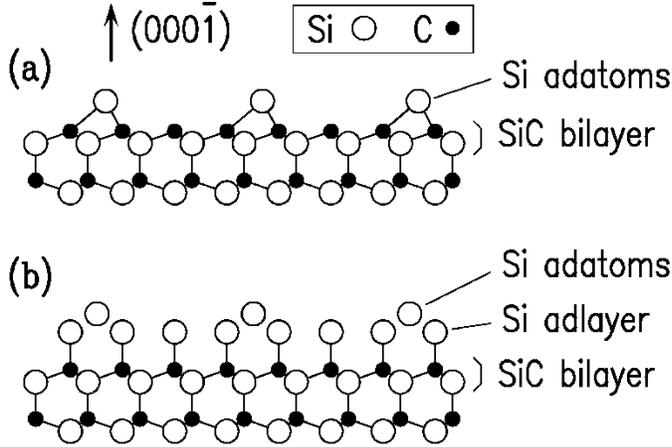

FIG 1. Possible terminations of the SiC(000$\bar{1}$) surface: (a) full SiC bilayer together with <1 monolayer of Si adatoms atop the bilayer, and (b) full bilayer plus a monolayer (adlayer) of Si plus additional Si adatom on top of the adlayer. Structures of the type shown in (b) are referred to as "adatom on adlayer" (AOA) structures.

The goal of the present work is to understand why graphene formation on C-face SiC under these near-equilibrium conditions (in disilane) appears to be so much different compared to when it is formed in high-vacuum conditions. Much of our work deals with reconstructions of C-face surface *in the absence* of graphene but nevertheless still under carbon rich conditions, i.e. heated to temperatures just below the point at which graphene forms. Figure 1 provides an introduction to two types of structures that we will consider: one with less than a monolayer (ML = 1 atom per SiC{0001} 1×1 surface unit cell = 12.1 atoms/nm$^2$) of Si adatoms, and the other with more than a ML. For wurtzite {0001} or zinc-blende {111} directions, planes of atoms in the bulk crystal form *bilayers*, as shown in Fig. 1. A natural way to form a surface is to preserve the bilayer at the surface, as in Fig. 1(a), such that the number of broken bonds is minimized. Although that type of surface termination does indeed occur in most cases, a few semiconductor surfaces reconstruct so as to split a surface bilayer,[26,27,28,29,30] as in Fig. 1(b); we refer to these structures as adatom-on-adlayer (AOA) structures. For the case of SiC(000$\bar{1}$) under C-rich conditions, most previously discussed surface structures are of the type shown in Fig. 1(a),[30,31,32,33] although one notable AOA structure of the type shown in Fig. 1(b) has been proposed.[34] In the present work, we find that it is actually AOA structures that are the energetically preferred ones, with or without overlying graphene, so long as the surfaces are formed at temperatures above ~400 K (which is true in all experimental cases).

Following a description of our methods in Section II, in Section III we demonstrate experimentally that a layer of Si atoms, consisting of ~1.3 monolayers (ML), exists between the graphene and the terminating SiC bilayer of the C-face SiC when the graphene is formed in disilane. This result is in contrast to the situation when graphene is formed in vacuum, when only ~0.55 ML of Si occurs between the graphene and the SiC bilayer (as evidenced by the dominant 3×3 interface reconstruction).[13,33] In Section IV, utilizing *ab initio* theory, we find that there are two energetically stable situations for Si terminating the C-face surface, one with ~0.55 ML of excess Si and the other with ~1.3 ML of excess Si. The former is stable for temperatures below about 400 K and the latter is stable for temperatures above that, with this different behavior arising from the effects of vibrational free energy. In Section V, we argue that the presence of the ~1.3 ML of excess Si accounts for at least some of the low-energy electron diffraction (LEED) results obtained from reconstructions of the surfaces/interfaces that we prepare in disilane (although additional work is needed to fully understand all of the observed reconstructions). We further argue that prior results for graphene formation in vacuum, although performed at temperatures >1000 K, were significantly Si deficient (i.e. under saturated) so that the resulting interface structures turned out to correspond to low-temperature and/or nonequilibrium ones.

The Si layer that we find to exist between the graphene and the SiC is important, not only in terms of its influence on the surface/interface structure, but also regarding the graphene formation. We argue that this layer serves a useful purpose, since subsequent oxidation of the layer (e.g. when samples are removed from the furnace or vacuum system in which they are formed) conveniently produces decoupling of the graphene from the SiC.[14,22]

## II. Experimental and Theoretical Methods

We form our graphene on the C-face of nominally on-axis 6H-SiC or 4H-SiC wafers (with no apparent differences between results for the two types of wafers) in a custom-built preparation chamber with an adjoining ultra-high vacuum chamber for low-energy electron diffraction (LEED) observation.[24,35] To remove polishing damage, the samples are first heated in either 1 atm of hydrogen at ~1600 °C for 3 min or 5×10$^{-5}$ Torr of disilane at 850 °C for 5 min, after which the surface display a 1×1 LEED pattern. Samples are then heated to a given temperature between 1150 and 1350 °C, and disilane is introduced to a pressure between 10$^{-6}$ and 10$^{-4}$ Torr (with most studies performed at 5×10$^{-5}$ Torr), with these conditions maintained for 5 min. Upon completion of the heating, the sample heater is turned off, requiring a few seconds to turn the potentiometer controlling the current completely to zero. Immediately after that the leak valve controlling the disilane pressure is turned off, requiring ~1 s.

We employ *ab initio* density functional theory (DFT) for thermodynamic computations,[36] utilizing both the Vienna Ab initio Simulation Package (VASP)[37,38] and the FHI-aims all-electron code[39,40] (results from the two methods, when identical structures are considered, agree within a few meV). All computations employ the Perdew-Burke-Enzerhof (PBE)[41] generalized gradient approximation (GGA) for a density functional, supplemented with van der Waals (vdW) interactions,[42] and with dipole corrections included according to the method of Neugebauer and Scheffler.[43] We set the plane-wave energy cutoff to 500 eV, and choose Γ-centered k-point grids with in-plane spacing of 1/40 Å$^{-1}$ or finer. Slabs consisting of six 3C-SiC bilayers with cubic lattice constant of 4.364 Å, (111)-oriented, and a graphene lattice constant of 2.463 Å are utilized (the difference in stacking order between cubic (111) planes and hexagonal (0001) planes is not expected to significantly affect energetic ordering of the various surface structures,[31] and since experimentally we do not know on which plane of the 4H or 6H crystals our surfaces occur, it is convenient simply to employ the 3C crystal structure in the theory). The bottom Si atoms in the slab are hydrogen terminated, and all the atoms in each structure considered are fully relaxed via conjugate gradients while holding the perpendicular and in-plane lattice constants fixed.

First-principles thermodynamics are employed,[44] with the temperature-dependent surface free energy of a given structure relative to that of a bare slab computed according to

$$\Delta\gamma(T) = \frac{1}{A}(E_{\text{struc}} - E_{\text{bare}} + F_{\text{struc}}(T) - F_{\text{bare}}(T) - \Delta N_{\text{Si}}\mu_{\text{Si}}(T) - \Delta N_{\text{C}}\mu_{\text{C}}(T) - \Delta N_{\text{H}}\mu_{\text{H}}(T)) \quad (1)$$

where $E_{\text{bare}}$ is the total energy of the bare SiC slab, $E_{\text{struc}}$ is the total energy of the surface structure after relaxation, $\Delta N_{\text{Si}}$, $\Delta N_{\text{C}}$, and $\Delta N_{\text{H}}$ denote the number of additional Si, C and H atoms, respectively, on the surface relative to the bare slab, and $\mu_{\text{Si}}(T)$, $\mu_{\text{C}}(T)$, and $\mu_{\text{H}}(T)$ are chemical potentials of Si, C and H atoms. The terms $F_{\text{bare}}(T)$ and $F_{\text{struc}}(T)$ are the vibrational free energies of the bare slab and of the surface structure, respectively. In thermal equilibrium we have

$$\mu_{\text{Si}}(T) + \mu_{\text{C}}(T) = E_{\text{SiC}} + F_{\text{SiC}}(T) \quad (2)$$

where $E_{\text{SiC}}$ is the internal energy per formula unit of bulk SiC, and with $F_{\text{SiC}}(T)$ being its vibrational free energy. Limits on $\mu_{\text{C}}$ and $\mu_{\text{Si}}$ are determined by the bulk phases, $\mu_{\text{C}}(T) \leq E_{\text{C}} + F_{\text{C}}(T)$ and $\mu_{\text{Si}}(T) \leq E_{\text{Si}} + F_{\text{Si}}(T)$, where $E_{\text{C}}$ and $E_{\text{Si}}$ are the internal energies per atom of Si and C atoms in bulk silicon and carbon, and with $F_{\text{C}}(T)$ and $F_{\text{Si}}(T)$ being their respective vibrational free energies. Using Eq. (2) to eliminate $\mu_{\text{Si}}$, we find the limits on $\mu_{C}$,

$$E_{\text{SiC}} - E_{\text{Si}} - E_{\text{C}} + F_{\text{SiC}}(T) - F_{\text{Si}}(T) \leq \mu_{\text{C}}(T) - E_{\text{C}} \leq F_{\text{C}}(T) \quad (3)$$

where we employ $E_{\text{C}}$ as a reference for $\mu_{\text{C}}(T)$. We use diamond-cubic silicon as the silicon bulk phase, graphite as the carbon bulk, and 3C SiC as the silicon carbide bulk, yielding $E_{\text{SiC}} - E_{\text{Si}} - E_{\text{C}} = -0.505$ eV. For $\mu_{\text{H}}(T)$, we list its values relative to $E_{\text{H}} = E_{\text{H}_2}^{\text{DFT}}/2$, where $E_{\text{H}_2}^{\text{DFT}}$ is the DFT-computed energy of the $H_2$ molecule.

The vibrational free energy terms $F_{\text{Si}}(T)$, $F_{\text{C}}(T)$, and $F_{\text{SiC}}(T)$ are all computed *ab initio*. Specifically, we calculate interatomic force constants using density functional perturbation theory in a supercell, evaluate the dynamical matrix at a dense set of wavevectors throughout the Brillouin zone, and diagonalize to obtain vibrational frequencies. Vibrational free energies are then evaluated from

$$F(T) = k_B T \int d\omega\, D(\omega)\, \ln[2\, \sinh(\hbar\omega/2k_B T)] \quad (4)$$

where $D(\omega)$ is the density of vibrational modes. For $F_{\text{struc}}$ and $F_{\text{bare}}$, these are computed using so-called Einstein modes, obtained from the *ab initio* computations by displacing a single atom while holding all other atoms fixed. $D(\omega)$ in these cases is given simply of a delta-function, $\delta(\omega)$, at the mode frequency.

The strategy that we employed in our structural search is as follows: We focused initially on the 2×2 AOA model suggested by Hibino et al.[34] However, when we tested that model using *ab initio* theory, its energy was found to be significantly higher than that of several other (non-2×2) structures of the SiC(000$\bar{1}$), hence casting doubt on this identification. We therefore conducted a search over all previously suggested SiC(000$\bar{1}$) models, plus variations thereof, seeking 2×2 or 4×4 models with energy lower than that of any other model (with any size unit cell), in the C-rich limit. One 4×4 AOA structure was identified at this stage that had energy lower than nearly all other models, but nevertheless this energy was still higher than that of the recently proposed 3×3 structure of Kloppenburg et al.[33] Hence, we turned to consider the possible role of H on the surface. However, despite a search through many models with 2×2, 4×4, and other unit cell sizes, we were never able to obtain energies lower than that of a simple H-terminated SiC bilayer or of the 3×3 structure of Kloppenburg et al. together with additional H termination. We therefore returned to structures without H, focusing on AOA models. A close examination of the results of Kloppenburg et al. for 2×2 cells led us to the realization that a 2×2 AOA model of the type proposed by Hoshino et al.[34] actually possesses additional distortions (implicit in the results of Kloppenburg et al.[33]) that significantly lowers its energy. We then examined many additional AOA models, fully considering all possible distortions of each and also including their vibrational free energies. More than 100 structural models were tested in total; results for the models with lowest energies are provided in Section IV, with additional results provided in the Supplemental Material.

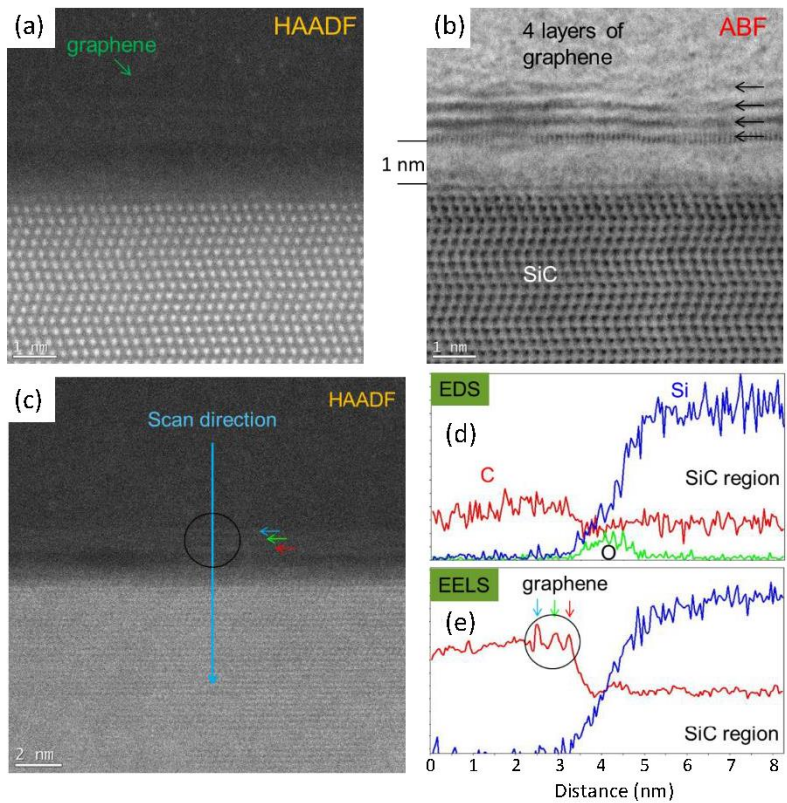

FIG 2. (a) High-angle annular dark-field (HAADF) imaging and (b) annular bright-field (ABF) imaging of graphene on C-face SiC. Four layers of graphene are observed for this region of the sample surface. These 4 layers show very low contrast in the HAADF image whereas they have clear contrast in the ABF image, consistent with the low atomic number of C compared with Si. (c) Another region of the sample, now displaying 3 layers of graphene. (d) EDS and (e) EELS measurements along the line indicated in (c).

## III. Experimental results

As described in prior work, when graphene is formed on C-face SiC under a disilane pressure of about 5×10⁻⁵ Torr, a characteristic $(\sqrt{43}\times\sqrt{43})$-$R\pm7.6°$ surface reconstruction occurs together with diffraction streaks associated with graphene.[14,22] Graphene formed in this manner is found to have considerably larger grain size than for graphene formed on the C-face in vacuum (2 μm vs. 50 nm grains).[14,23,24] We have studied one such sample by cross-sectional scanning transmission electron microscopy (STEM), with results shown in Fig. 2. This sample was found to be covered with 2 – 4 layers of graphene, depending on surface region. An amorphous layer with thickness of about 1.0 nm was found to be present between the top layer of SiC and the bottom layer of graphene, as seen in Figs. 1(a) – 1(c). Averaging over multiple surface regions, the thickness of this layer was found to be 0.9±0.1 nm. Energy dispersive X-ray spectroscopy (EDS) measurements, Figure 2(d), indicate that the interfacial layer is silicon oxide, since increases in the Si and the O spectra are observed as the scan enters the interfacial layer, while the C signal decreases. Similar results are obtained by the electron energy loss spectroscopy (EELS) measurements, Fig. 2(e), with the EELS line scan explicitly showing the 3 graphene layers present at this surface region.

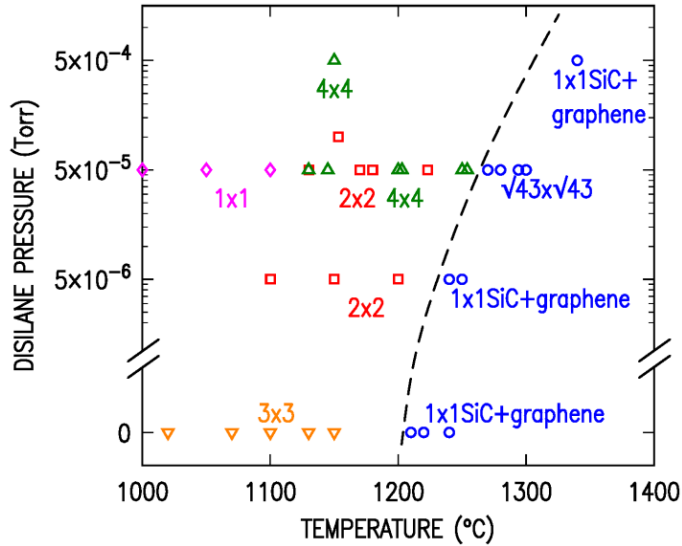

FIG 3. Overview of experimental LEED results, showing the symmetry of the observed patterns. Each data point represents a surface prepared by 5 minutes of heating at the temperatures and disilane pressures indicated. LEED patterns to the right of the dashed line contain graphene spots (indicating the presence of graphene on the surface), while those to the left of the dashed line do not contain graphene spots. The "0" disilane pressure (vertical axis) indicates experimental results obtained in vacuum, with no disilane introduced.

The oxygen present in the observed $SiO_x$ interface layer presumably arises from the several-month time that the sample sat in air between production and the STEM study. (Based on extensive experience with our preparation system we are confident that no significant oxygen is present at the surface or interface *during* graphene formation.[14,15,22,24,45] Oxidation produces a C-face surface with characteristic $(\sqrt{3}\times\sqrt{3})$-$R30°$ LEED pattern;[46] we observe that if we intentionally expose our surface to oxygen, but we *never* see if otherwise, even after extended heating of the surfaces).[24] In terms of Si content, using a mass density of $SiO_x$ of 1.8 g/cm$^3$ (with x≈1.5)[47] along with the 0.9±0.1 nm thickness, the Si content at the interface is found to correspond to 1.55±0.17 ML. We emphasize that no heating of the samples was performed between the graphene preparation and the STEM observation, so that the observed Si at the interface must have formed, or been present, during the graphene growth. We thus surmise that $1.4 – 1.7$ ML of Si exists at the interface between graphene and C-face SiC, when the graphene is formed in ~5×10$^{-5}$ Torr of disilane.

Returning to the characteristic $(\sqrt{43}\times\sqrt{43})$-$R\pm7.6°$ surface reconstruction that is observed on such samples, we have performed additional LEED studies on surfaces that are formed at temperatures just below those where graphene forms. Figure 3 shows a summary of our LEED observations, showing the symmetry of observed patterns as a function of temperature and disilane pressure. The LEED patterns are displayed in Fig. 4. In the absence of any disilane, an intense 3×3 pattern is observed on our C-face SiC surface after heating in ultra-high vacuum (UHV) to a temperature of ~1000°C, in agreement with prior works.[13,14,15,24,48] At higher temperatures, graphene forms on the C-face surface by the well-known mechanism of preferential sublimation of Si atoms.[49] Most importantly, this preferential sublimation of Si atoms means that, as a function of temperature, the surface is becoming more C rich (i.e. high values of C chemical potential). When disilane is introduced at pressure of about 5×10$^{-6}$ Torr or higher, the situation changes dramatically. We then observe only 2×2 or 4×4 LEED patterns [Figs. 3(b) and 3(c)] at temperatures below the onset of graphene formation (this graphene formation temperature also increases substantially as the disilane pressure increases, as expected). Samples showing 2×2 patterns and those with 4×4 patterns were prepared using nominally the same procedures; at disilane pressure near 5×10⁻⁵ Torr, there appears to be some subtle (not well understood) difference in surface conditions that determines whether one or the other of these structures is obtained (the 1×1 phase seen at lower temperatures than the 2×2/4×4 is presumably a disordered, kinetically limited stucture). As we further increase the temperature to form graphene, we observe the $(\sqrt{43}\times\sqrt{43})$-$R\pm7.6°$ reconstruction for disilane pressures near 5×10⁻⁵ Torr, as previously reported.[14] In addition to the $(\sqrt{43}\times\sqrt{43})$-$R\pm7.6°$ spots (which we identify in the present work to arise from an AOA structure terminating the C-face SiC), these patterns also contain weak graphene spots/streaks centered typically at $\pm7°$ relative to the SiC (1,0) spots. For the case of disilane pressures of 5×10⁻⁶ Torr or 5×10⁻⁴ Torr, we obtain a "1×1SiC+graphene" pattern, which contains streaks at the graphene spot locations (again, at $\pm7°$ relative to the SiC (1,0) spots), along with the underlying SiC spots; this pattern is indicative of thin, multi-domain graphene covering the surface.

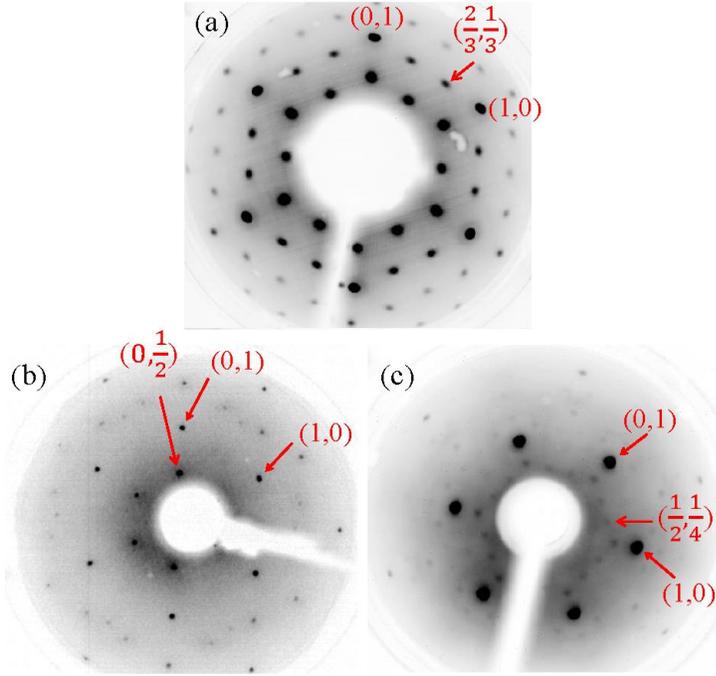

FIG 4. LEED patterns of (a) 3×3, (a) 2×2, (c) 4×4 surfaces. Pattern (a) was obtained from a sample heated at 1070 °C for 10 min in vacuum (without any disilane), whereas patterns (b) and (c) were obtained from samples prepared under nominally identical conditions: heating in 5×10⁻⁵ Torr disilane at 1180 °C for 5 min. Patterns (a) and (c) were acquired at an electron energy of 100 eV, and pattern (b) at 96 eV.

Regarding the 4×4 pattern that we observe, a unit cell with this size has not been previously reported on SiC(000$\bar{1}$) surfaces, to our knowledge. However, for the 2×2 pattern, there are two well-known examples of that in prior work, arising from the (2×2)$_C$ and (2×2)$_{Si}$ surfaces, so named since the latter is more Si-rich than the former.[50] The former surface structure has been definitely identified by Seubert et al.[51] in a LEED intensity vs. energy (I-V) analysis to consist of a single Si adatom per 2×2 cell residing on a SiC bilayer (as in Fig. 1(a)). The latter structure has been proposed by Hoshino et al.[34] on the basis of medium energy ion scattering and photoelectron spectroscopy to also arise from a single Si adatom per 2×2 cell, but with the adatom in this case residing on a Si adlayer atop the SiC bilayer (as in Fig. 1(b)). We have measured LEED I-V spectra from our 2×2 and 4×4 surfaces, as shown in Fig. S1 of the Supplemental Material. The spectra from the two surfaces are very similar, indicating some close connection between their respective structures. In any case, the 2×2 spectra show poor agreement with those reported for the (2×2)$_C$ surface, indicating that our 2×2 surface does not have that structure. For the case of (2×2)$_{Si}$, we are not aware of previously reported LEED I-V spectra. Nevertheless, comparing our LEED spot intensities of Fig. 4(b) (at 96 eV) with those previously reported in Ref. [52] for (2×2)+Si (at 95 eV), we find quite good agreement – the {1/2,0} spot intensities as large as those for {1,0} spots, but the {1/2,1/2} spots are absent (e.g. one of these, if visible, would lie directly between (0,1) and (1,0) in Fig. 4(b)). Moreover, the formation procedure of our surface and the (2×2)$_{Si}$ ones of the prior works are similar.[50,34] Hence, as a starting hypothesis for our structure, we utilize the model of Hoshino et al. consisting of a single Si adatom per 2×2 unit cell atop a Si adlayer.[34]

## IV. Theoretical results

The goal of our theoretical computation is to identify the structures that give rise to our observed 2×2, 4×4, and ($\sqrt{43}\times\sqrt{43}$)-$R\pm7.6°$ LEED patterns. We first discuss the C-face surface in the absence of any graphene on it and neglecting any possible role of H atoms. We show in Fig. 5 the total free energy of various surface structures, as a function of the coverage of Si adatoms for each structure. We focus on C-rich conditions, i.e. when there is zero energy associated with the formation of graphite/graphene on the surface (aside from possible interaction of the graphite/graphene with the underlying SiC surface atoms). The notation used for structural models provides the dimensions of the unit cell, followed by the number of additional adatoms over and above a terminating SiC bilayer on the surface. Fig. 6 shows the atomic arrangements for each of the labelled structures of Fig. 5. No structures containing C atoms are found to be energetically favorable, in agreement with prior works.[30,31,32,33] The high energies of such structures occur because of the relatively strong bonding of C within graphite, so that C atoms in SiC surface reconstructions would always prefer to be in graphite (or graphene), or in the SiC bulk, rather than in some surface reconstruction.

Figures 5 and 6 display results for two models with Si adatom coverage of < 1 ML. At a coverage of 0.25 ML is the well-known (2×2)+Si model, Fig. 6(b), denoted (2×2)$_C$ in past work and consisting of a Si adatom on a three-fold hollow site of the SiC bilayer, i.e. with one remaining C "rest atom" of the bilayer that is not bonded to the adatom.[50,51] At a coverage of $5/9 \approx 0.556$ ML is the novel (3×3)+5Si model recently proposed by Kloppenburg et al., consisting of 5 Si adatoms on the 3×3 cell, Fig. 6(a), with 3 of the adatoms residing in bridging sites of the bilayer and the remaining 2 adatoms then adopting three-fold coordinated sites between these bridging atoms. There is strong evidence that this model provides the explanation for the 3×3 LEED pattern that is commonly observed on surfaces prepared by heating in vacuum (Figs. 3 and 4(a)).[13,33,53] The energy of this structure is lower (under C-rich conditions) than that of any other SiC(000$\bar{1}$) surface structure known prior to the Kloppenburg et al. work.

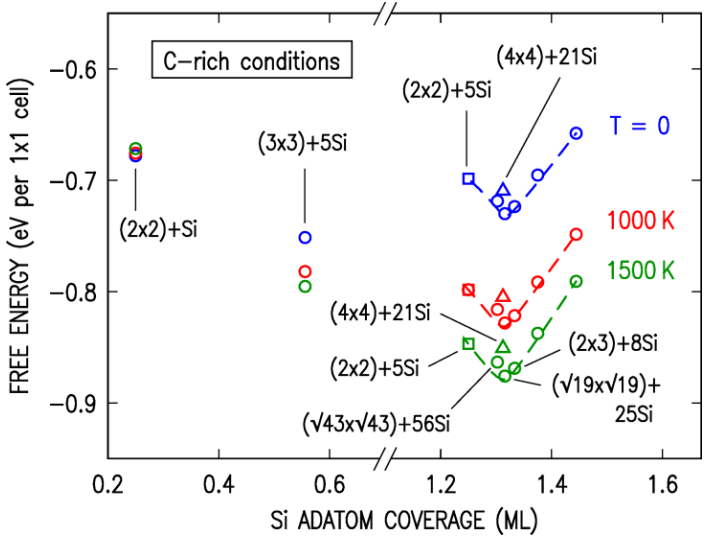

FIG 5. Total energy, including vibrational free energies, for select surface structures on SiC($\overline{1}\,\overline{1}\,\overline{1}$) (intended to model SiC(000$\overline{1}$) surfaces), as a function of the coverage of Si adatoms. Results are shown in the C-rich limit, and for temperatures of 0, 1000, and 1500 K. The dashed lines for coverages near 1.3 ML are drawn to match the minimum-energy models there, for each temperature.

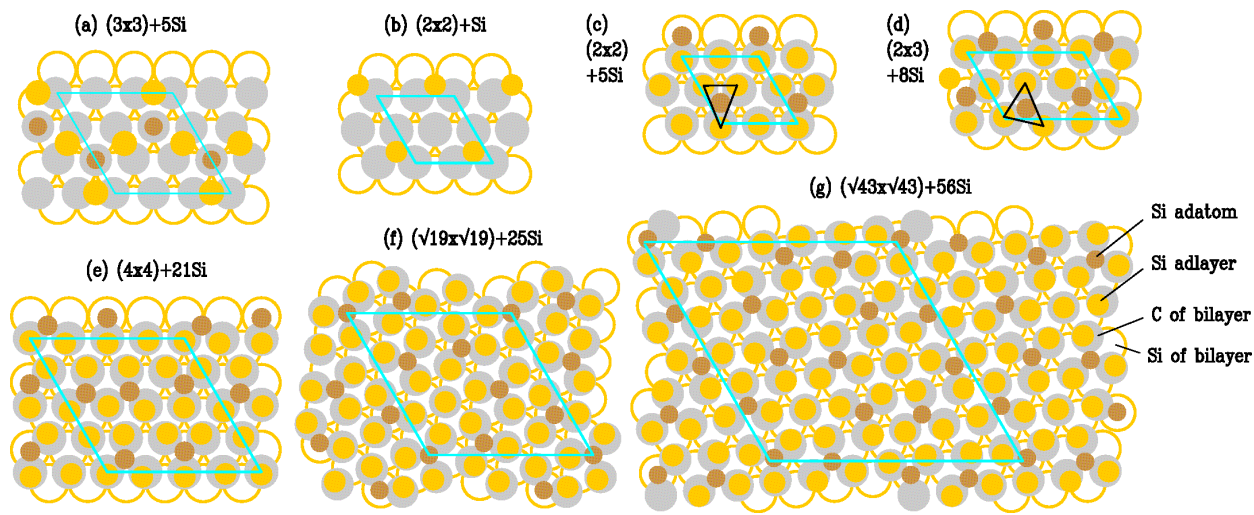

FIG 6. Schematic views of the surface structures whose energies are labeled in Fig. 5. The surface arrangements sit on a SiC bilayer, with Si atoms (open orange circles) at the bottom of the bilayer and C atoms (filled gray circles) at the top. Above the bilayer are Si adatoms (solid orange circles), which in many cases form a complete adlayer. Above these adatoms, for certain structures, are additional Si adatoms (brown filled circles). Unit cells are shown by a light blue trapezoid. The black triangle in (c) indicates a biaxial distortion in the location of the 3 orange adlayer atoms at its corners, whereas the triangle in (d) shows a predominantly twisting distortion.

All other models in Figs. 5 and 6 have >1 ML of Si adatoms, i.e. being AOA models. The adatoms atop the adlayer can reside in sites that are directly *on top* of Si atoms located 3 layers below in the SiC bilayer or at three-fold *hollow sites* that are between those atoms. Distortions of the adlayer atoms around the adatoms are found to be quite important in reducing the total energies. The (2×2)+5Si model shown in Fig. 6(b) consists of a Si adatom on a Si adlayer; it is essentially the model proposed by Hishino et al.,[34] although with the adatom being in an on-top (rather than hollow) site and with a significant biaxial distortion as shown in Fig. 6(c) (the isosceles triangle there is distorted away from an equilateral one). This biaxial distortion reduces the energy by 48 meV per (1×1) unit cell relative to a structure with C$_{3v}$ symmetry, as first deduced by Kloppenburg et al.[33] The structure shown in Fig. 6(c), with its adatom in an on-top site (relative to Si atoms 3 layers below), has an energy that is 52 meV/(1×1) lower than when the adatom is in a hollow site.

The other AOA models in Figs. 5 and 6 are all new, previously unreported ones. The (2×3)+8Si model of Fig. 6(d) has equal numbers of adatoms in on-top and hollow sites, and it displays a significant "twisting" type distortion (as known from prior models[30,54]). as shown in the figure. The ($\sqrt{19}\times\sqrt{19}$)+25Si and ($\sqrt{43}\times\sqrt{43}$)+56Si models also display significant twisting type distortions, whereas the (4×4)+21Si model shows a combination of biaxial and twisting distortions. These three models have unequal numbers of adatoms in on-top and hollow sites. In these cases, there is a "complementary" structure with reversed numbers of adatoms in on-top and hollow sites; usually a given structure has energy within a few meV/(1×1) of its complement, although the (2×2)+5Si structure mentioned in the prior paragraph is an exception to this trend.

The results of Fig. 5 include the effects of vibrational free energy, and hence they vary depending on the temperature. Most importantly, we find that, at low temperatures, the lowest energy surface structure is the (3×3)+5Si one at 0.556 ML adatom coverage,[33] whereas at elevated temperatures, the lowest energy structures are the ones near 1.3 ML adatom coverage. Thus, as the temperature increases to above about 380 K, a surface layer of Si atoms with coverage of ~1.3 ML is seen to form, under equilibrium conditions. The stability of this layer, for temperatures above ~380 K, originates from two effects (of comparable magnitude) arising from vibrational free energy. The first is the inherent temperature dependence of the C-rich limit. This limit is given by a value of carbon chemical potential of $\mu_C(T) = E_C + F_C(T)$. In Fig. 7, we display $F_C(T)$ as obtained from an *ab initio* computation, shown along with $F_{Si}(T)$ and $F_{SiC}(T)$. We see that $F_C(T)$ varies between $+0.171$ eV at zero temperature down to a value of $-0.100$ eV at 1500 K, with the latter temperature corresponding to what we use in our experiments. Hence, $\mu_C(T)$ in the C-rich limit will vary by this same amount, so that at elevated temperature, structures that are more Si-rich will be favored.

The second effect of vibrational free energy is a shift in energies of *each* of the model structures due to their *individual* vibrational entropies, i.e. the term $F_{struc}(T) - F_{bare}(T)$ in Eq. (1). In evaluating this term, it is important to realize that there is an additional aspect of Eq. (1) that acts to partially offset the term. From Eq. (2) we have $\mu_{Si}(T) = -\mu_C(T) + E_{SiC} + F_{SiC}(T)$, and substituting that into $-\Delta N_{Si}\mu_{Si}(T)$ from Eq. (1) yields $-\Delta N_{Si}(-\mu_C(T) + E_{SiC} + F_{SiC}(T))$. Therefore, for the structures we are considering with nonzero $\Delta N_{Si}$, they will have a temperature-dependent contribution to their total energy of $-\Delta N_{Si}F_{SiC}(T)$. Together with the $F_{struc}(T) - F_{bare}(T)$ term, we then must evaluate $F_{struc}(T) - F_{bare}(T) - \Delta N_{Si}F_{SiC}(T)$. In essence, $F_{struc}(T) - F_{bare}(T)$ produces a reduction in total energy due to the vibrational entropy of the additional Si atoms on the surface, but this reduction is partially offset by the $-\Delta N_{Si}F_{SiC}(T)$ contribution which is the *negative* of the vibrational free energy of the same number of both Si and C atoms in SiC.

We evaluate $F_{SiC}(T)$ using *ab initio* methods, Fig. 7, and in principle, $F_{struc}(T) - F(T)$ can be evaluated in a similar manner. However, due to the large size of the unit cell for some of our surface structures, it is not computationally feasible to do so. Hence, we adopt an approximate method for estimating the $F_{struc}(T) - F_{bare}(T)$ term, utilizing so-called *Einstein modes*,[36] obtained from the *ab initio* computations by displacing a single atom while holding all other atoms fixed. Table I lists a few of these energies, both for bulk materials and for surface structures (i.e. computed using our 6-bilayer slabs). With these energies, computed for relevant atoms in each structure, the vibrational free energy is obtained from Eq. (4) with $D(\omega)$ being a delta-function for each mode.

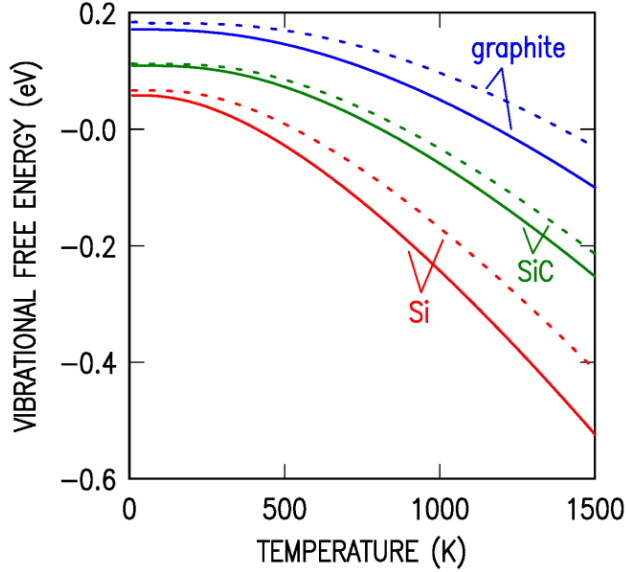

FIG 7. Vibrational free energies for graphite, Si, and 3C-SiC, with all results shown per atom (for SiC, the result shown is 1/2 the free energy per Si+C unit). Solid lines show free energies obtained using a complete spectrum of vibrational modes, whereas dashed lines show results using only a single Einstein mode for Si and graphite, and two such modes for SiC. All mode energies are obtained from *ab initio* computations.

Use of the Einstein modes for evaluating the structure-specific contribution to the vibrational free energy is rather approximate. For example, Fig. 7 shows *ab initio* free energies for bulk structures using a full spectrum of vibrational modes compared with those obtained using Einstein modes. The significant errors encountered by use of the latter is apparent, with the true free energies being significantly more negative than those obtained from the Einstein modes. Considering this error, a more realistic computation of the AOA structures of Fig. 5 would cause them to shift downwards in energy, as a function of increasing temperature, *faster* than pictured there. However, aside from this aspect, the differences in energies for neighboring AOA structures near the minimum of the total energy curve, i.e., with coverage of 1.25 − 1.35 ML, will be scarcely affected. Hence, this approximate treatment of the vibrational modes has no significant impact on our final results.

**Table I**. Energies of Einstein modes, obtained by displacing a specified atom in a structure (inequivalent atoms are specified by numbers). When three energies are listed, the first two modes correspond to motion in the plane of the surface, and the third is perpendicular to the surface.

| structure | energy (meV) |
|---|---|
| Si in bulk SiC (8-atom cell) | 60.8 |
| C in bulk SiC (8-atom cell) | 88.7 |
| Si in bulk Si (8-atom cell) | 44.3 |
| (3×3)+5Si, bridging atom | 32.1, 32.6, 46.6 |
| (3×3)+5Si, adatom 1 | 28.6, 28.6,  47.1 |
| (3×3)+5Si, adatom 2 | 31.2, 31.2, 48.0 |
| (2×2)+5Si, adlayer atom 1 (rest atom) | 20.0, 28.0, 44.0 |
| (2×2)+5Si, adlayer atom 2 (not rest atom) | 25.8, 32.3, 49.8 |
| (2×2)+5Si, topmost adatom | 20.0, 33.0, 34.1 |

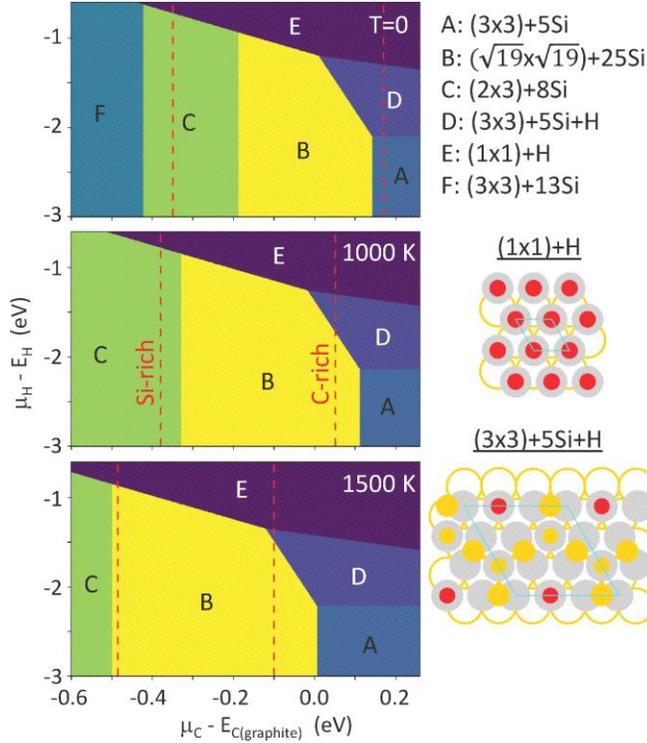

FIG 8. Computed phase diagrams, showing regions of minimum free energy for various SiC(000$\bar{1}$) surface structures, at temperatures of 0, 1000, and 1500 K. Vibrational free energy is included, leading to the temperature dependence of the Si-rich and C-rich limits (dashed lines) as well as the temperature-dependent shifts in the positions of boundaries between phases. Structures A, B, and C are shown in Figs. 6(a), 6(f), and 6(d), respectively. Structures D and E are shown in the schematic views, using the same color scheme as in Fig. 6 and with red filled circles representing H (these two structures are the only ones in the phase diagram that contain any H). Structure F is the one proposed in Ref. [30].

Turning now to the presence of graphene formation over the C-face reconstructed surfaces, as was first pointed out by Neugebauer and Northrup, graphene is expected to interact only relatively weakly with underlying atoms.[55] To estimate this interaction energy, we have performed computations on a few select reconstructions of the C-face surface that are covered in graphene, as presented in the Supplemental Information. We find, not surprisingly, that the strain energy of the graphene (i.e. when it is forced to fit specified supercells of the SiC) can be quite significant. However, when this energy is subtracted from the total, then the remaining interaction between the graphene and the underlying atoms (Si adatoms) is found to be about $-34$ meV/(1×1). As can be seen by examination of Fig. 5, this energy is relatively small compared to the energetics associated with the formation of a Si layer on the C-face surface. Hence, the graphene is found to play only a relatively small role in the Si adlayer formation, but nevertheless it could well have some influence on the precise arrangement of adatoms on top of the adlayer.

To complete our discussion of the energetics of C-face SiC surfaces, we must consider the possible presence of H atoms on the surface. We have undertaken an extensive series of *ab initio* computations for structural models that include H, as described in Fig. S3. Resulting phase diagrams showing the minimum-energy structures as a function of the chemical potentials of H and C are shown in Fig. 8. We find only two surface structures containing H that are minimum-energy, stable ones, for any physically realizable values of H and C chemical potential: One structure is the (1×1)-H surface, consisting of a H-terminated SiC bilayer. The other is formed by having a H atom terminate the single C rest atom (i.e. not bonded to an adatom) that is present in the (3×3)+5Si model, thereby forming (3×3)+5Si+H.

The primary reason that no other structures involving H are stable is that the Si-H bond is quite weak compared to the H-H bond. Therefore, for AOA models, H atoms will always prefer to stay bonded in the form of $H_2$, rather than bond to surface Si atoms. For the case of H bonding to surface C atoms, our computations do not reveal any minimum-energy, stable structures aside from the (1×1)+H and (3×3)+5Si+H surfaces shown in Fig. 8. For example, considering a SiC bilayer simply containing Si adatoms (of any coverage) directly bonded to the bilayer, and with remaining C rest atoms being bonded by H, we find that such surfaces are *always* higher in energy than the (1×1)+H or (3×3)+5Si+H surfaces. We have also studied surfaces covered with a nearly complete adlayer of Si atoms, e.g. with one Si atom of the adlayer missing and the C rest atom thus formed being terminated by a H atom. The energies of such surfaces are found, in some cases, to be not too much higher than those of other, non-H-containing structures. Nevertheless, such surface with partial adlayers are never found to form minimum-energy structures. Since the only minimum-energy surfaces containing H are found to be (1×1)+H and (3×3)+5Si+H, and our experiments in disilane do *not* reveal any surfaces with (1×1) or (3×3) periodicity, we conclude that the surfaces in our experiments are inconsistent with structures that contain H. This conclusion is consistent with the general tendency of H termination to yield unreconstructed or minimally reconstructed surfaces.[56]

## V. Discussion

The main conclusion of our work, based on the *ab initio* thermodynamics of Fig. 5, is that a Si layer (with ~1.3 ML coverage) forms on the SiC C-face surface for temperatures above about 400 K. This conclusion provides an explanation for the experimental results of Fig. 2, although we note that the amount of Si detected between the graphene and the SiC there, ranging from 1.4 – 1.7 ML depending on surface location, is somewhat higher than the ~1.3 ML from the theory. Perhaps there are other surface structures that we have not considered that have even more Si, although it must be remembered that such structures would have to be minimum energy ones even in the C-rich limit (and this limit does not favor structures with even more Si). Alternatively, perhaps for samples such as in Fig. 2, there could be some excess (i.e. over-saturated) Si at the interface due to the Si subliming from the SiC/graphene interface, i.e. as the graphene grows thicker.

Further confirmation of the presence of the Si layer on the surface can be obtained from an identification of the detailed structures giving rise to the LEED patterns that we obtain (Fig. 3). We believe that there is ample evidence that our 2×2 pattern arises from the (2×2)$_{Si}$ surface structure, as identified in prior studies and also shown in Fig. 6(c) by (2×2)+5Si.[33,34] Theoretically, this structure (including its distortion away from $C_{3v}$ symmetry) was first identified by Kloppenburg et al.,[33] and considering the very large number of models investigated in that work, it is very unlikely that any other 2×2 model can be found that has total energy lower than this (2×2)+5Si structure. Experimentally, this same surface was found by Hoshino et al. to contain nearly a full Si adlayer, plus one additional Si adatom per 2×2 cell on the adlayer.[34] It should be noted that the simulations performed in that work relied on a structure with $C_{3v}$ symmetry, i.e. neglecting the biaxial distortion of the structure. Energetically, we find that this distortion reduces the energy by 48 meV/(1×1), so it is very important. More to the point, if the Hoshino et al. analysis was redone using the distorted structure, we expect that they would arrive at the on-top geometry for the adatom as being in best agreement with experiment, since this geometry is substantially favored (compared to the hollow site) in terms of its total energy, as discussed in Section IV.

Nevertheless, their determination of the presence of the Si adlayer for the structures appears to be definitive,[34] independent of the biaxial distortion.

Despite this apparent agreement with the present and past experiments with the theoretically predicted (2×2)+5Si structure, we find in Fig. 5 that the free energy of this structure is slightly *higher* than that of other AOA models, such as the ($\sqrt{19}$×$\sqrt{19}$ )+25Si structures. We suggest two possible reasons for this discrepancy. The first is that we have neglected configurational free energy in our analysis. The (2×2)+5Si structure has 3 possible orientations for its distortion (with a barrier of 48 meV/(1×1) separating these energy minima). Thus, we expect a configurational contribution of $-(k_B T \ln 3)/(2×2)$, or $-36$ meV/(1×1) at 1500 K. This is a substantial effect, on the scale of Fig. 5. However, we find that part of this energy reduction is offset by configurational contributions to other AOA models, for which we can move individual adatoms in the models to form many additional configurations. These other arrangements have slightly higher energies than the respective ground states, but we account for those using a partition function, and from that we obtain free energy changes of typically $-20$ meV/(1×1) at 1500 K for this effect. A more detailed analysis is required to thoroughly evaluate these configurational free energies, but in any case we are confident that the resulting reductions in energy (relative to what is shown in Fig. 5) will be greatest for the (2×2)+5Si structure.

The second possible reason why the (2×2)+5Si structure is *not* found to be an overall energy minimum in Fig. 5, but it *is* observed in experiment, has to do with the density of adatoms on the experimentally prepared surfaces. It is possible that this density might be affected by details of the preparation. At the temperatures of ~1500 K used for the surface preparation, it is likely that a disordered, lattice gas system of adatoms is present on the surface at these temperatures, similar to what occurs for Si adatoms on the Si(111) surface.[57,58] Following the heating, the current to the heater is shut off and the temperature of the sample drops rapidly, at an estimated rate of several hundred K per second judging by the change in brightness of the heater.[35] As the temperature drops, at some point the surface structure will become kinetically frozen in. It is difficult to accurately estimate that temperature, although we know from our studies of preparation of the 3×3 surface in vacuum (Fig. 3, plus additional data off the left-hand side of that plot) that a temperature of 1200 – 1300 K is required to reliably form that reconstruction (i.e. to enable significant bond breaking and atomic motion on the surface); this value possibly can be taken as a lower bound for the temperature at which the surface structure freezes in. We feel that it is possible that this temperature is reached for our surfaces in a sufficiently short time, i.e. upon cooling from 1500 K, so that the adatom density may not have time to equilibrate. Experiments with additional annealing over extended times at temperatures of, say, 800 – 1000 K (and under an appropriate disilane pressure), could conceivably reveal other surface structures, such as the ($\sqrt{19}$×$\sqrt{19}$ )+25Si arrangement.

For the 4×4 pattern, the lowest energy 4×4 model that we have found is the (4×4)+21Si structure shown in Fig. 6(e), but we see in Fig. 5 that this structure has free energy that is significantly higher than that of other models, such as the ($\sqrt{19}$×$\sqrt{19}$ )+25Si arrangement. At this time, we cannot assign the structure that gives rise to our observed 4×4 pattern. Our estimates of configurational free energies, mentioned above, produce results that do not seem to significantly reduce the (4×4)+21Si free energy relative to that for ($\sqrt{19}$×$\sqrt{19}$ )+25Si. We have studied >25 different 4×4 AOA models, but nevertheless, perhaps additional ones need to be investigated in order to find a lower energy structure. Alternatively, possibly additional forms of surface disorder (e.g. including a dilute mixture of 4×4 cells in a surrounding 2×2 surface) need to be considered. In any case, as discussed in Section III, the fact that our 2×2 and 4×4 LEED patterns are produced using the same surface preparation procedures, and the LEED I-V intensities of these two patterns are in good agreement, provides evidence that the 4×4 structure, like the (2×2)+5Si one, consists of an adlayer of Si with some additional adatoms on top of the adlayer.

Moving to our observed $(\sqrt{43}\times\sqrt{43})$-$R\pm7.6°$ LEED pattern, we see from Figs. 5 and 6(g) that there is a structure with this symmetry, with total energy only 12 meV/(1×1) cell higher than that of our minimum-energy $(\sqrt{19}\times\sqrt{19})$+25Si structure. Experimentally, the $(\sqrt{43}\times\sqrt{43})$-$R\pm7.6°$ is only observed when graphene covers the surface, whereas the 2×2 or 4×4 arrangements are seen in the absence of graphene. Possibly the graphene layer *stabilizes* the $(\sqrt{43}\times\sqrt{43})$+56Si adatom arrangement in some way; in Section IV we described small, but significant, interaction energies that can occur between adatoms and overlying graphene. We have not performed such computations specifically for graphene on top of the $(\sqrt{43}\times\sqrt{43})$+56Si structure, due to the very large commensurate unit cell that forms between the graphene and the interface structure. It should also be noted that the we have investigated only a few structures with the relatively large unit cell size of $\sqrt{43}\times\sqrt{43}$, so any identification of the experimental pattern with the structure of Fig. 6(g) must be considered as quite tentative.

It is apparent that in past studies of reconstructions on SiC($000\bar{1}$), nonequilibrium surface structures are commonly formed. The well-known (2×2)$_C$ reconstruction on this surface is firmly established to consist of a single Si adatom per 2×2 cell, residing directly on the SiC bilayer (Fig. 6(b)).[51] It is clear from Fig. 5, where this structure is denoted (2×2)+Si, that it is *not* an energy minimum (for any temperature). Rather, its formation must be a result of under-saturation of the surface Si concentration (although it should be noted that this reconstruction may well form predominantly on graphene-covered surfaces[30]). Similarly, the predominant 3×3 pattern observed on the surface (Fig. 4(a)) has been recently identified to arise from the (3×3)+5Si structure[33] (Fig. 6(a)). As seen in Fig. 5, this structure is indeed an energy minimum for temperatures less than about 400 K. However, experimentally, the surface is formed by annealing at temperatures much higher than that. Again, under-saturation of the surface Si content apparently occurs. Even for the (2×2)+5Si surface, formed both in our work and prior works in a Si-rich environment (from disilane or a Si flux),[51,34] it is apparent from the above discussion that it may also form in slightly under-saturated conditions.

Our conclusion regarding the Si layer on the surface, or between the graphene and the SiC surface, provides explanation for other experimental observations as well. For example, in our past work, we prepared graphene on C-face SiC samples using a disilane environment, then removed the sample from the vacuum chamber and exposed it to air (or pure oxygen), followed by annealing in high vacuum at 1000 °C for several minutes. Such samples then displayed an intense $(\sqrt{3}\times\sqrt{3})$-$R30°$ LEED pattern;[22] LEED I-V analysis revealed that this pattern originated from a *silicate* layer below the graphene, i.e. with the silicate having the $(\sqrt{3}\times\sqrt{3})$-$R30°$ structure as first elucidated (in the absence of graphene) by Starke et al.[46] The silicate layer itself contains 5/3 ML of Si, so its occurrence is totally consistent with the presence of the Si adlayer below the graphene prior to oxygen exposure. (Indeed, from the presence of the silicate alone, one could perhaps have concluded that a Si adlayer was present during the graphene formation. However, we avoided that conclusion in our prior work because of the several-minute 1000-°C annealing step that was found to be necessary in order to form the silicate, since that annealing step could, conceivably, liberate Si from the SiC crystal itself.[22])

Our conclusion is also consistent with, and provides a partial explanation for, the report of Wang et al. concerning intercalation of Si at the interface between graphene and C-face SiC.[59] They found that, for temperatures > 1020 °C, silicon that has been pre-deposited on graphene on C-face SiC will readily move to the SiC/graphene interface, and they intercalated as much as 6 ML of Si in this manner. Those authors noted that the reason for the intercalation was not known, since, in equilibrium, the excess Si was expected to bond with the C atoms from the graphene to make SiC. Our determination of Si-rich surface reconstructions in the C-rich limit (for temperatures above 380 K) explains the intercalation, at least the first ~1.3 ML of it. As to the mechanism for forming the interface concentration of Si over that amount, this is not easy to understand and provides an interesting topic for future study.

Nicotra et al.[60] have reported TEM characterization of the C-face SiC/graphene interface, revealing an interface layer quite similar to our results of Fig. 2. Excess Si is found directly above the SiC, with some apparent oxidation (which could have occurred during growth or post growth). The major new result of our present work is the demonstration that the presence of the excess Si, i.e. a Si adlayer plus additional Si adatoms, is an *equilibrium property* of the SiC surface, both with or without graphene present above the Si AOA structure. The data of Nicotra et al., as well as our data of Fig. 2, also indicates the possible presence of excess C in this interface layer, located near the overlying graphene. As discussed in Section IV, it is energetically unfavorable for C to mix with Si in the AOA structure (the C atoms would prefer to form graphene). We also note that there are no known, thermodynamically stable, ternary compounds containing Si, C, and O.[61] Nevertheless, during graphitization of the C-face SiC, it is possible that some excess C is located in the Si AOA layer below the graphene, i.e., due to kinetic limitations that prevent it from immediately forming the thermodynamically stable graphene phase.

Finally, we discuss the prior work of de Heer and others, mentioned in Section I, in which exceptionally good electronic transport behavior has been obtained for graphene formed on C-face SiC, under confined conditions.[18,19,20,21,25] Based on the results of the present work, we feel that it is possible that a Si adlayer exists between the graphene and the C-face surface (the latter terminated by a SiC bilayer) in their experiments, given that their growth is conducted at about 1200 °C under conditions that are very near to equilibrium.[18] However, this conclusion is not consistent with their reported results of surface/interface characterization.[62] Using crystal truncation rod (CTR) analysis in surface x-ray diffraction, the best-fit model of Hass et al. contains a top-most SiC bilayer with only ~0.7 ML of both the Si and C atoms and with the C layer being corrugated.[62] Hence, no Si adlayer is present in that model. However, we note that structural analysis of CTR data is highly dependent on the assumed model of the interface. For example, if we replace the corrugated C layer of this model (with its total density of 0.74 ML) by Si atoms, at a density of $0.74/2.33 = 0.32$ ML which would make their x-ray signal comparable to that from the corrugated C layer,[62] we would then arrive at an AOA-type model for the interface. Additionally, it is possible that oxidation of this interface occurs during growth (since their furnace is not under UHV conditions[18]), in which case a more complicated model with additional layers would have to be considered. We also note that the corrugated C layer in the model of Hass et al.,[62] if it bonds directly to graphene as they suggest, has a very unfavorable total energy for the reasons described in Ref. [30] (also summarized in the second to last paragraph of the Supplemental Material). Further work is necessary to more completely understand the similarities and differences between the graphene-covered C-face SiC surface from our work compared to those of de Heer and co-workers.

## VII. Summary

In summary, we have observed experimentally the presence of a Si-rich interface (>1 ML of Si) between C-face SiC and graphene formed by thermal decomposition of the SiC. To explain the presence of this excess Si, we propose Si AOA structures, which are found to have minimum free energy under C-rich conditions, so long as nonzero temperatures are considered. These structures are stabilized by a combination of vibrational and configurational free energies.

We emphasize that the presence of the Si adlayer terminating the SiC($000\bar{1}$) surface along with the associated AOA structures are *equilibrium* properties of the surface, existing over the full range of chemical potentials varying between Si-rich to C-rich conditions. We distinguish between several situations that can occur during graphene formation on C-face SiC: (i) With sufficient overpressure of Si, nearly equilibrium conditions will prevail with a Si adlayer present between the SiC and graphene. We have demonstrated this to be the case in our work, and we consider it likely to also apply during growth in confinement-controlled conditions.[18,19,20,21] (ii) When heating the surface in vacuum, it is likely that the Si adatoms and adlayer will sublimate from the surface quite readily at even moderate temperatures, hence leading to nonequilibrium surface structures (for $T > 400$ K) such as the observed (2×2)+Si and (3×3)+5Si arrangements that lack the adlayer.[13,53] (iii) Oxidation of the surface during graphene formation is a critical issue, since even trace amounts of oxygen will lead to the formation of the energetically stable ($\sqrt{3} \times \sqrt{3}$)-$R30°$ silicate structure on the surface which is found to inhibit graphene formation.[14,15,22,24,45,46] Inhomogeneous graphene formation can result from the presence of this layer.[15] However, perhaps with sufficient Si overpressure, and with near-equilibrium conditions (albeit including oxygen), uniform growth can be established. The reported growths under confinement-controlled conditions have *not* been performed under the pure, UHV conditions necessary to eliminate oxidation,[18] and hence it is possible that a silicate-type layer exists in those cases. It should be noted that the silicate layer itself decomposes in a vacuum environment for temperatures above 1200 °C, but it has been found to be stable in an 1-atm argon environment for temperature up to 1600 °C.[15] Its possible presence under confined conditions remains to be investigated.

Inclusion of vibrational free energies was found to be essential in our work, to achieve the level of agreement between experiment and theory that we obtain. Similar effects may occur on other surfaces as well. For example, we point out the prior work of Ga adlayers on N-face and Ga-face GaN{0001} surfaces, i.e. the GaN($000\bar{1}$) and (0001) surfaces, respectively, for which vibrational free energy was *not* included.[26,27,28,29] The N-face results are very analogous to those presented in the present work, in that Ga adlayers are found to form on the surface even under N-rich conditions (due both to the relatively large size of Ga compared to N, enabling the adlayers to form, and to the energetic stability of $N_2$ molecules, so that a N-terminated surface is relatively unstable).[26] Although vibrational free energy was not considered in that case, there doesn't appear to be any discrepancy between experiment and theory, i.e., under N-rich conditions there is *always* a Ga adlayer present on the surface.[26] However, for the Ga-face, there does appear to be some discrepancy between theory and experiment, under Ga-rich conditions. The experimental observations clearly demonstrate the existence of multiple adlayers of Ga on those surfaces, but structural models that have the same symmetry as the experiments consistently produce *ab initio* energies that are not minimum ones.[27,29] The experimental evidence for the Ga adlayers is sufficiently strong so that little or no doubt exists as to their presence, and hence there must be some reason that the *ab initio* results do not produce minimum energies. We suggest that vibrational (and/or configurational) free energies may provide an explanation for that discrepancy between experiment and theory.

## Acknowledgements

This work was supported by the U.S. National Science Foundation, grants DMR-1205275 and 1809145, and by the Department of Energy, grant SC-0014506. MJK was supported in part by Global Research and Development Center Program (2018K1A4A3A01064272) and Brain Pool Program (2019H1D3A2A01061938) through the National Research Foundation of Korea (NRF) funded by the Ministry of Science and ICT. Discussions with Phillip First are gratefully acknowledged.

**Formation of Graphene atop a Si adlayer on the C-face of SiC**

Jun Li,[1] Qingxiao Wang,[2] Guowei He,[1] Michael Widom,[1] Lydia Nemec,[3] Volker Blum,[4] Moon Kim,[2] Patrick Rinke,[3,5] and Randall M. Feenstra[1]

[1]Dept. Physics, Carnegie Mellon University, Pittsburgh, PA 15213, USA
[2]Dept. Materials Science and Engineering, The University of Texas at Dallas, Richardson, TX 75080, USA
[3]Fritz-Haber-Institut der Max-Planck-Gesellschaft, D-14195, Berlin, Germany
[4]Dept. Mechanical Engineering and Materials Science, Duke University, Durham, NC 27708 USA
[5]Dept. Applied Physics, Aalto University, P.O. Box 11100, Aalto FI-00076, Finland

## I. Additional Experimental Results

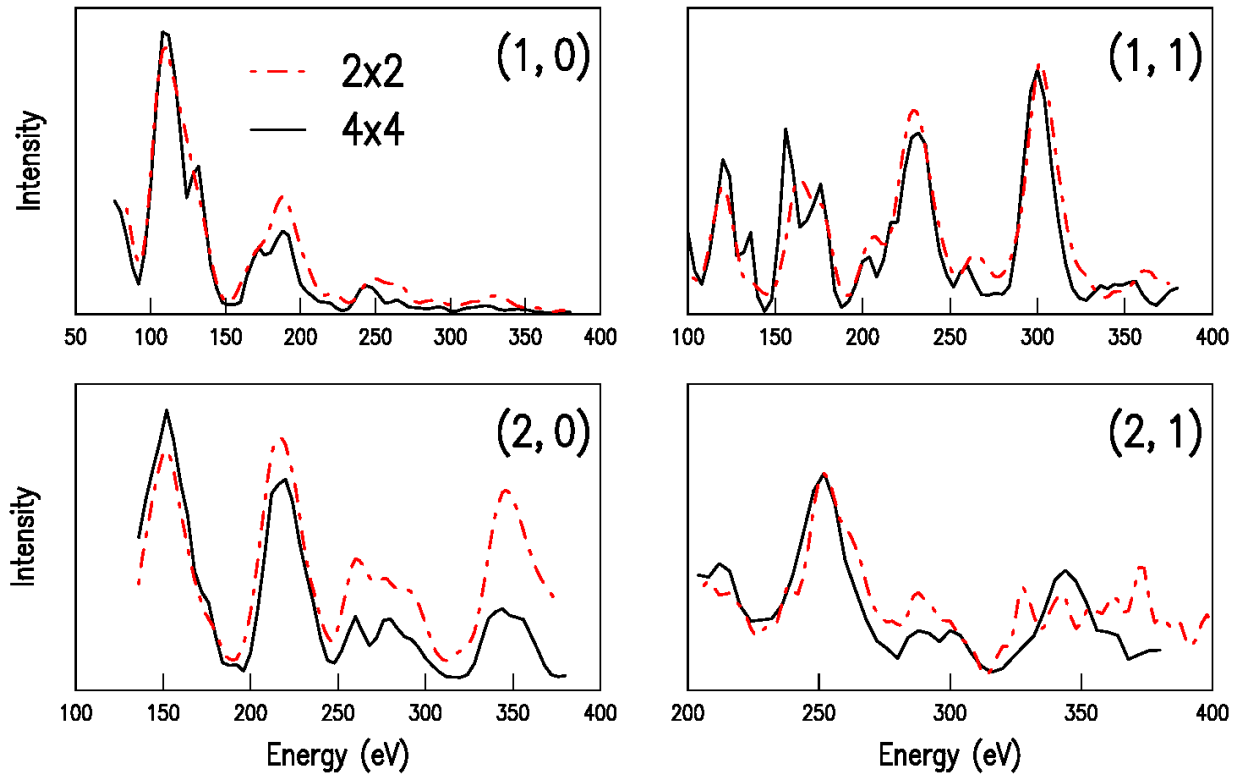

FIG S1.  LEED I-V spectra of the 4×4 surface compared to that of the 2×2 surface. Results are obtained by averaging the intensities of all equivalent spots of a given type that are visible in the LEED patterns, focusing here on integer-order spots (which the two surfaces have in common). The I-V curves for the two surfaces are seen to be very similar, implying that the structures of the two surface are likely to be closely related. The two surfaces were prepared under nominally the same conditions: heating in 5×10⁻⁵ Torr of disilane at 1160 °C for 5 min.

## II. Additional Theoretical Results
### A) Surface Structures without Hydrogen

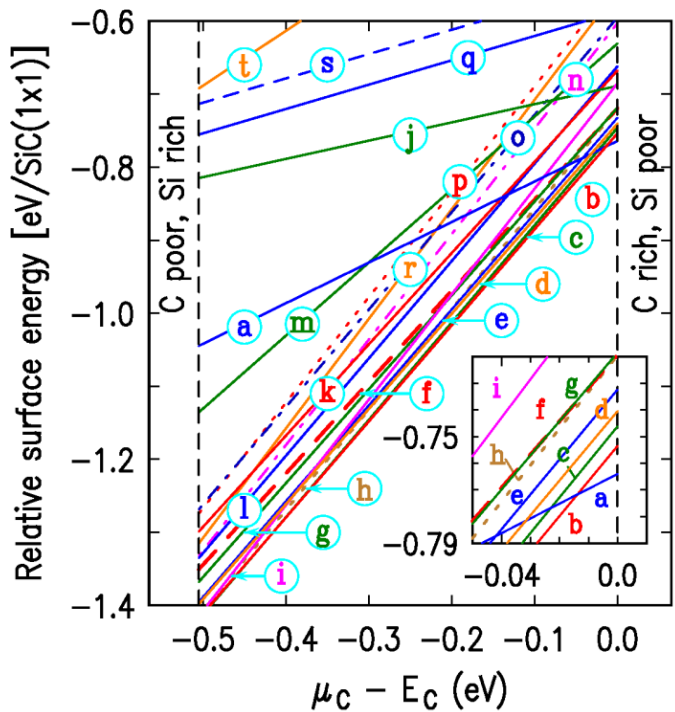

FIG S2. Ab-initio computational results for various structures, without H and not including vibrational effects. Surface energies are plotted as a function of the C chemical potential within the allowed ranges (given by Si in the diamond structure and C in the graphite structure). Where multiple structures have been considered with given unit cell size and numbers of adatoms, generally only the lowest energy one is displayed. Notation for the various surface structures lists the number of atoms of various types, including information as to their bonding site and height relative to the surface. For example, (3×3)+5Si(3b,2t) has 5 Si adatoms per 3×3 cell, with 3 of those in bridging (b) sites and 2 in on-top (t) sites,[1] as seen in Fig. 6(a). The site specifications in this case are relative to the C atoms of the uppermost SiC bilayer. The 't' adatoms are located in a plane that is slightly above that of the 'b' adatoms, hence separated from them in the specification by a comma. As another example, (3×3)+13Si(ml,3t,t) has 13 Si adatoms, 9 of them in a monolayer-thick (ml) adlayer, 3 above the adlayer in on-top sites relative to the C atoms 2 layers below, and 1 adatom that is above those

t) (2x2)+4Si(ml)+C(h)
s) (√3x√3)+Si(t)
r) (2x2)+6Si(ml,2t)
q) (√3x√3)+Si(h)
p) (3x3)+13Si(ml,3t,h)
o) (√3x√3)+4Si(ml,t)
n) (3x3)+13Si(ml,2h+2t)
m) (√3x√3)+3Si(ml)
l) (3x3)+12Si(h+2t)
k) (2x2)+5Si(ml,h)

j) (2x2)+Si(h)
i) (3x3)+13Si(ml,3t,t)
h) (4x4)+22Si(ml,2h+4t)
g) (√7x√7)+9Si(ml,h+t)
f) (2x2)+5Si(ml,t)
e) (4x4)+21Si(ml,2h+3t)
d) (√43x√43)+56Si(ml,7h+6t)
c) (2x3)+8Si(ml,h+t)
b) (√19x√19)+25Si(ml,3h+3t)
a) (3x3)+5Si(3b,2t)

other 3 adatoms, in an on-top site relative to the C atoms 3 layers below.[2] A further example is (4×4)+21Si(2h+3t), as shown in Fig. 6(f). This model has 21 Si adatoms, 16 of them in a ml and 5 in a layer above that; 2 of those 5 are in 'h' sites relative to the C atoms 2 layer below and the other 3 are in 'h' sites relative to those C atoms. (As is apparent from these examples, the layer that is employed as a reference for specifying bonding sites varies from one set of adatoms to another. The site specification is not intended to provide a *complete* definition of the structures, i.e. since it also doesn't include specific information on *which* b, h, or t sites in a cell are occupied; rather it is intended only to provide some partial means of distinguishing the models). Only a single structure is shown that contains C adatoms, namely, (2×2)+4Si(ml)+C(h) with a single C adatom atop a Si adlayer. Its energy, along with the energies of all structures containing C adatoms, is relatively high, as explained in the main text. Structures (a) – (f) are shown in Fig. 6 (a), (g), (e), (h), (f), and (d), respectively, with structure (j) shown in Fig. 6(b). Structures (h) and (i) have energies shown by the two unlabelled points at coverages of 22/16 and 13/9, respectively, in Fig. 5.

## B) Surface Structures with Hydrogen

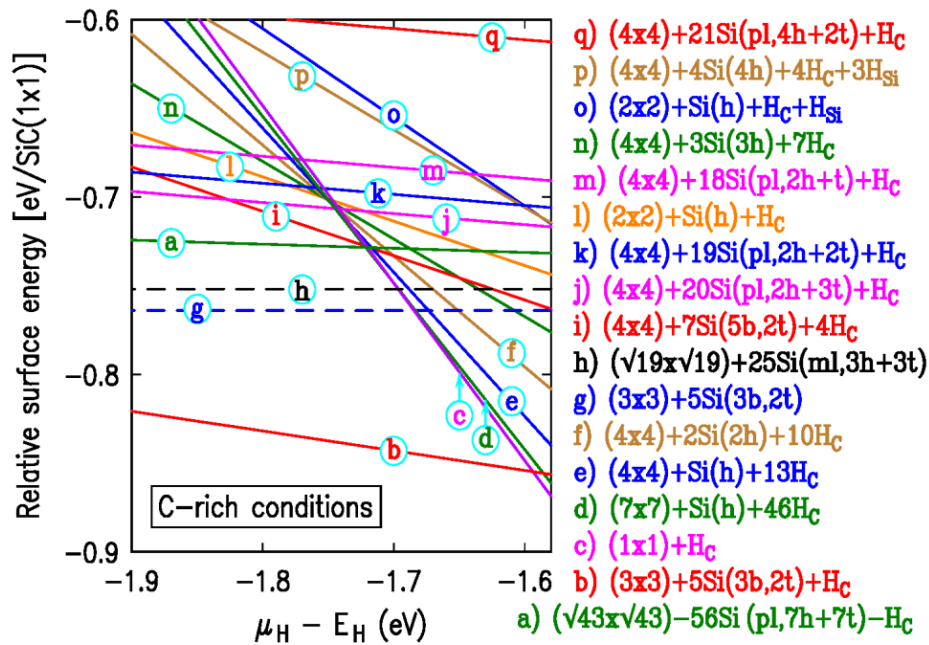

q) **(4x4)+21Si(pl,4h+2t)+H$_C$**
p) **(4x4)+4Si(4h)+4H$_C$+3H$_{Si}$**
o) **(2x2)+Si(h)+H$_C$+H$_{Si}$**
n) **(4x4)+3Si(3h)+7H$_C$**
m) **(4x4)+18Si(pl,2h+t)+H$_C$**
l) **(2x2)+Si(h)+H$_C$**
k) **(4x4)+19Si(pl,2h+2t)+H$_C$**
j) **(4x4)+20Si(pl,2h+3t)+H$_C$**
i) **(4x4)+7Si(5b,2t)+4H$_C$**
h) **($\sqrt{19}$x$\sqrt{19}$)+25Si(ml,3h+3t)**
g) **(3x3)+5Si(3b,2t)**
f) **(4x4)+2Si(2h)+10H$_C$**
e) **(4x4)+Si(h)+13H$_C$**
d) **(7x7)+Si(h)+46H$_C$**
c) **(1x1)+H$_C$**
b) **(3x3)+5Si(3b,2t)+H$_C$**
a) **($\sqrt{43}$x$\sqrt{43}$)−56Si(pl,7h+7t)−H$_C$**

FIG S3. Ab-initio computational results for various structures, including the possible presence of H but not including vibrational effects. The number of H atoms is indicated in the notation (subscript 'C' denotes bonding to C atoms), and 'pl' refers to a partial adlayer of Si adatoms atop the uppermost SiC bilayer. This partial adlayer is missing one Si adatom (a rest atom) per unit cell, with the corresponding C atom of the bilayer then being bonded to a H atom. In general, it is found that the only energetically favorable location for H to bond is to a C atom, and the only energetically favorable location of C atoms is in the SiC bilayer. A model for structure (a) is shown in Fig. S4. Over the range of H chemical potential shown, the lowest energy model is (3×3)+5Si(3b,2t)+H$_C$, consisting simply of the (3×3)+5Si(3b,2t) structure[1] along with the C rest atom in that structure being terminated by H. The model (4×4)+7Si(5b,2t)+4H$_C$ is a variation on this (3×3)+5Si(3b,2t)+H$_C$ structure, in which additional bridging adatoms enable the overall arrangement to be spaced out to a 4×4 cell and the additional C rest atoms thus formed are all terminated by H. Structures such as (7×7)+Si(h)+46H$_C$ consist of a single Si adatom on the uppermost SiC bilayer, with all other C surface atoms being terminated by H.

### ($\sqrt{43}$x$\sqrt{43}$)+56Si (pl,7h+7t)+H$_C$

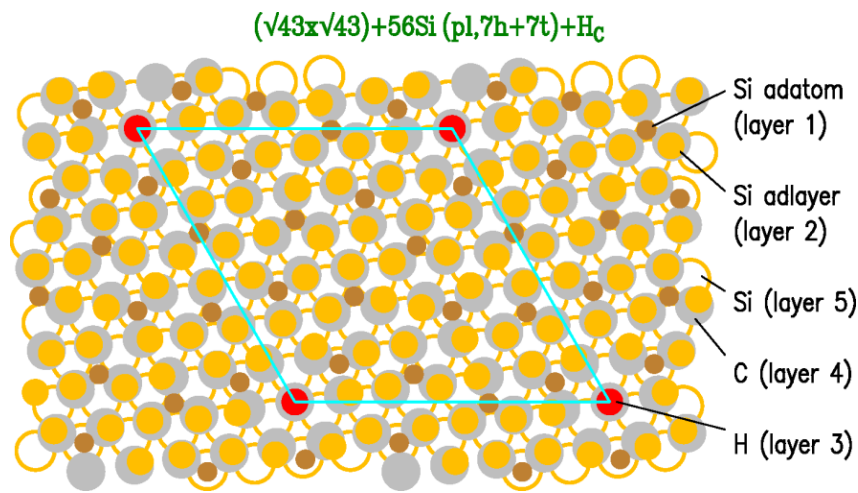

Si adatom (layer 1)
Si adlayer (layer 2)
Si (layer 5)
C (layer 4)
H (layer 3)

FIG S4. Model for structure (a) of Fig. S3. Hydrogen atoms are bonded to C atoms of the SiC bilayer, with a single H atom per unit cell. The energy of this model always is *higher* than that of other models, such as (b), (c), (g), or (h) of Fig. S3 (depending on H chemical potential), but not too much higher.

## C) Surface Structures with Graphene

We consider here the situation when a layer of graphene is included on the surface (and vibrational free energies are neglected). We have considered only a few such structures, two of which are shown in Fig. S5. The structures shown there are not intended to be complete models of the graphene on SiC system, but rather, they are hypothetical structures designed to elucidate the role of strain in the graphene layer. Graphene is placed on a H-terminated SiC(000$\bar{1}$) surface, using either 8×8 graphene cells (128 C atoms) on a ($\sqrt{43}$×$\sqrt{43}$)+43H$_C$ SiC surface or 13×13 graphene cells (338 C atoms) on a ($6\sqrt{3}$×$6\sqrt{3}$)+108H$_C$ structure. The latter is well known from prior studies of Si-face SiC as being a good, low-strain fit between graphene and SiC,[3,4,5] whereas the former is one that was proposed in our prior work as a possible fit between the graphene and the SiC for the case of C-face SiC, i.e. as a possible explanation for the observed $\sqrt{43}$×$\sqrt{43}$ periodicity. As seen in Fig. S5, the total energy of the first model is considerably higher than that of the second, by 0.13 eV per 1×1 SiC cell.

We attribute this energy difference between the two models to strain of the graphene layer for the 8×8 graphene on the ($\sqrt{43}$×$\sqrt{43}$)+43H$_C$ structure. Table S1 lists some lattice constants for graphene and SiC, $a_C$ and $a_{SiC}$ respectively. If we consider $n_C$×$n_C$ unit cells of graphene strained to fit onto $n_{SiC}$×$n_{SiC}$ cell of SiC, the lattice mismatch will be $f = (n_{SiC}a_{SiC}/\sqrt{2} - n_C a_C)/n_C a_C$, where the surface lattice parameter for SiC is $\frac{a_{SiC}}{\sqrt{2}}$ with $a_{SiC}$ being the cubic, bulk value. The mismatch equals the in-plane strains, $\varepsilon \equiv \varepsilon_{xx} = \varepsilon_{yy}$, in the graphene. Evaluating the strains, in the upper part of Table S1 we list our theoretical lattice constants, 4.364 Å for 3C-SiC and 2.463 Å for graphite (both of which happen to agree well with the experimental values at room temperature).[6,7] With these values, the in-plane strain in the graphene for the $\sqrt{43}$×$\sqrt{43}$ structure is $\varepsilon = 0.0270$. Employing graphene elastic constants of $C_{11} + C_{12} = 414$ N/m and $2C_{111}/3 - C_{222}/3 + C_{112} = -1026$ N/m,[8] we obtain a biaxial strain energy change of $(C_{11} + C_{12})\varepsilon^2 + (2C_{111}/3 - C_{222}/3 + C_{112})\varepsilon^3 = 0.145$ eV/ SiC(1×1). This value is reasonably close to our 0.13 eV energy difference, thereby confirming our interpretation in terms of strain.

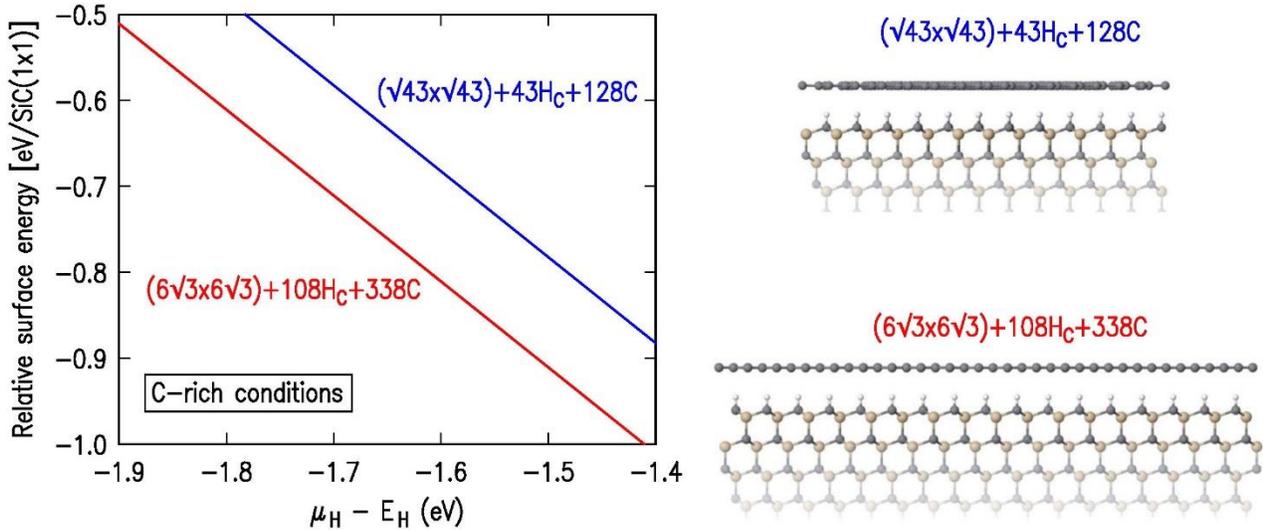

FIG S5. Total energy for surface structures on SiC($\overline{111}$) (intended to model SiC(000$\bar{1}$) surfaces), including graphene. Energies relative to the bulk-terminated (1×1) phase are plotted as a function of the H chemical potential, for C-rich conditions and without consideration of vibrational free energy. Side-view models for the structures are shown, with carbon atoms represented by dark gray balls, silicon by yellow balls, and hydrogen by light gray balls.

| $a_C$ | $a_{SiC}/\sqrt{2}$ | $n_C$ | $n_{SiC}$ | $f$ |
|---|---|---|---|---|
| $2.463^a$ | $3.086^a$ | 8 | $\sqrt{43}$ | 0.0270 |
| | | 13 | $6\sqrt{3}$ | 0.0016 |
| | | $\sqrt{57}$ | 6 | -0.0043 |
| $2.463^{a,b}$ | $3.097^b$ | 8 | $\sqrt{43}$ | 0.0306 |
| | | 13 | $6\sqrt{3}$ | 0.0052 |
| | | $\sqrt{57}$ | 6 | -0.0007 |

Table S1. Lattice misfit $f$ for various commensurate fits of graphene on SiC, as computed for given lattice parameters (Å) of the graphene and the SiC. Upper 3 rows show lattice parameters of *ab initio* theory, and lower rows show estimated values appropriate to a temperature of 1500 K.

[a] *ab initio* theory (this work)

[b] estimated at 1500 K from theory and/or experiment (Refs. [6,7,9,10])

This relatively large biaxial strain in a structure with 8×8 graphene unit cells matched to $\sqrt{43} \times \sqrt{43}$ SiC unit cells becomes even worse if we employ lattice constant appropriate to the preparation temperature of ~1500 K, as shown in the lower part of Table S1 (the lattice parameter of graphite at 1500 K is not well known, but it appears to be little different than the room temperature value and hence we employ our theoretical value in this case).[9,10] Indeed, re-examining the $(\sqrt{43} \times \sqrt{43})$-$R\pm7.6°$ experimental LEED patterns,[11] it is clear that the graphene does *not* have a 8×8 fit to the graphene, but rather its diffraction spot is a streak located at a wavevector that is 2.0±0.5% larger than that of $8/\sqrt{43}$ times the (1,0) SiC spot, and this streak is spread over angles of about 5 – 8° relative to the (1,0) SiC spot.

To better match this experimental diffraction result for the graphene, we consider a different match between graphene and SiC, namely, a $(\sqrt{57} \times \sqrt{57})$-$R\pm6.5°$ graphene cell fitting on a 6×6 SiC surface. This match is a relatively good one, as previously identified by Hass et al.,[3] and indeed Table I reveals a mismatch of only $\varepsilon = -0.0007$ at 1500 K. Additionally this type of model will yield diffraction spots for the graphene that are located at ±6.5° relative to the principal SiC directions, which is quite close to the experimental observations not only for the $(\sqrt{43} \times \sqrt{43})$-$R\pm7.6°$ pattern but also for other C-face graphene/SiC surface structures as well for which streaks centered at about ±7° relative to the principal SiC directions are commonly found.[12] We thus conclude that the graphene in our experiments forms on the SiC in this sort of structure (or with slight rotations about this structure, i.e. as evidenced by the streaks in the LEED patterns), i.e. a $(\sqrt{57} \times \sqrt{57})$-$R\pm6.5°$ graphene cell fitting on a 6×6 SiC cell. Of course, the actual periodicity of the SiC surface/interface is $(\sqrt{43} \times \sqrt{43})$-$R\pm7.6°$. Hence, in order to obtain a coincidence between these two SiC cells (and hence between the Si adatoms and the overlying graphene), it is necessary to create a very large SiC cell, with periodicity of $(6\sqrt{43} \times 6\sqrt{43})$-$R\pm 7.6°$.

We also comment on two structural models containing graphene *without* any H atoms, models (e) and (c) in Fig. 3 of Ref. [13], to illustrate the energetics involved. The first of these, a $(6\sqrt{3} \times 6\sqrt{3})$+338C structure containing 13×13 graphene cells (338 C atoms) placed on a $6\sqrt{3} \times 6\sqrt{3}$ surface unit cell of a bulk-terminated SiC(000$\bar{1}$) surface, is found to have a relatively high energy: $-0.16$ eV per 1×1 SiC relative to the bulk terminated surface, in the C-rich limit. In contrast, for the second model which is similar but with a $2 \times 2$ arrangement of Si adatoms (located on three-fold hollow site) terminating the SiC, i.e. $(6\sqrt{3} \times 6\sqrt{3})$+27Si(h)+338C, the energy is significantly lower: $-0.68$ eV per 1×1 SiC cell. This value is nearly the same as for the (2×2)+Si(h) model of Fig. S2 (same as model (b) in Fig. 3 of Ref. [13]), indicating very weak interaction between the graphene and the adatom-terminated surface. More significantly, the much higher energy of the $(6\sqrt{3} \times 6\sqrt{3})$+338C model reveals the unfavorability of that type of structure. Bonds form between the terminating C atoms of the SiC bilayer and the C atoms of the graphene in that model, as clearly revealed by the rumpling of the graphene (see Fig. 3 of Ref. [13]). However, whatever energy benefit arises from these bonds (e.g. by improved saturation of the C atoms of the SiC surface) is apparently largely outweighed by the energy cost of overall poorer bonding for the C atoms of the graphene layer itself. We have not computed additional structures containing graphene on an AOA layer on the C-face, but based on the results of Fig. 5 of the main text we can be confident that their energies will be less than that of the $(6\sqrt{3} \times 6\sqrt{3})$+27Si(h)+338C structure by $0.1 - 0.2$ eV/1×1 cell for temperatures in the range $500 - 1500$ K. In the C-rich limit the overlying graphene has little impact on the surface energetics, so long as the SiC surface is properly terminated with Si adatoms or a Si AOA structure (or H) such that it doesn't significantly bond to the graphene.

To achieve a more quantitative estimate of the interaction energy between graphene and an underlying adatom-terminated surface, we compare the energies of structures with and without a graphene layer on top, in the C-rich limit. One structure to consider is the $(6\sqrt{3} \times 6\sqrt{3})$+108H$_C$+338C model of Fig. S5, which is found to be 40 meV per 1×1 SiC unit cell higher than the H-terminated SiC bilayer surface, (1×1)+H$_C$. Another structure is the $(6\sqrt{3} \times 6\sqrt{3})$+27Si(h)+338C model just discussed, having graphene lying on top of a simple 2×2 adatom covered surface; the total energy of this surface is 6 meV/1×1 higher than without the graphene. The fact that, in both cases, we find *higher* energies with the graphene than without (in the C-rich limit) is not surprising, since the definition of C-rich limit is when zero energy is needed to form graphite, and we must not forget the van der Waals interaction between the graphene planes within the graphite. In any case, if we further take the difference between the two energy differences just stated, we arrive at an estimate for the interaction energy between graphene and adatoms of $-34$ meV/1×1 SiC cell for this 2×2 arrangement of adatoms.